\documentclass[twocolumn,showpacs,preprintnumbers,amsmath,amssymb,pra]{revtex4}
\usepackage{graphicx}
\usepackage{dcolumn}

\begin{document}
\title{Coherently prepared nondegenerate Y-shaped four-level  correlated emission laser: A source of tripartite entangled light}

\author{Sintayehu Tesfa} \email{sint_tesfa@yahoo.com}
\affiliation{Max Planck Institute for the Physics of Complex Systems, N$\ddot{o}$thnitzer Str. 38, 01187 Dresden, Germany\\
Physics Department, Dilla University, P. O. Box 419, Dilla, Ethiopia}

\date{\today}

\begin{abstract} A detailed derivation of the master equation  of the cavity radiation of a coherently prepared $Y$-shaped four-level correlated emission laser is presented. The outline of the procedures that can be employed in analytically solving the stochastic differential equations and the rate equations of various correlations are also provided. It is shown that coherently preparing the atoms in the upper two energy levels and the lower, initially, can lead to a genuine continuous variable tripartite entanglement. Moreover, preparing the atoms in the coherent superposition, other than the possible maximum or minimum, of the upper two energy levels, leaving the lower unpopulated, may lead to a similar observation. With the possibility of the atom at the intermediate energy level to take three different transition roots guided by the induced coherence, this system in general is found to encompass versatile options for practical utilization. In particular, coupling at least one of the dipole forbidden transitions by an external radiation is expected to enhance the degree of detectable entanglement.\end{abstract}

 \maketitle

 \section{INTRODUCTION}

In recent years, due to the relative simplicity and high efficiency in the generation, manipulation, and detection of optical continuous variable (CV) states \cite{rmp77513}, the corresponding entangled states have been successfully implemented in unconditional quantum teleportation \cite{prl80869,s282706}, quantum dense coding \cite{prl88047904}, quantum error correction \cite{np5541}, and universal quantum computation \cite{prl821784} among others. With the progress in the CV entanglement research, the generation of more than a bipartite entanglement has attracted much attention. A truly $N$-partite entangled state generated by a single-mode squeezed state and linear optics \cite{prl91080404} along with the generation of CV tripartite entanglement using cascaded nonlinear interaction in an optical cavity without linear optics have been theoretically investigated \cite{pra74042332} and also experimentally realized \cite{prl91080404,prl97140504}. Nevertheless, the structure of the entanglement  for the three-mode system is a bit more subtle than that for a bipartite case, wherein, different classes of entanglement are defined based on how the density matrix may be partitioned \cite{pra74063809}. The classifications range from fully inseparable, which means that the density matrix is not separable for any grouping of the modes (genuine tripartite entanglement), to fully separable, where the three modes are not entangled in any way. Despite the challenge, fortunately, there is a large number of works that are aimed at devising ways of detecting a tripartite entanglement \cite{pra64052303,pra67052315,pra74063809,pra75012311,pra81062322}. 

On the other hand, there has been enormous effort in studying quantum optical systems that are capable of generating tripartite entangled light \cite{pra72053805,pra74042332,pra74063809} that includes the nondegenerate parametric oscillation and six-wave mixing in the nonlinear medium. As an alternative, the idea of generating a tripartite entanglement from various schemes of four-level atomic systems via coherent superposition has been under study recently \cite{jpb41035501,jpb42165506,oc2821593,jpb43155506,jmo561607,pra82032322,epjd56247}.  Entangled photons from these systems are expected to have a potential applications in quantum memory \cite{n432482} and long distance quantum-communication \cite{prl855643}, since the low frequency and narrow linewidth of the light can enhance efficient coupling between photons and atomic memories in a quantum network \cite{prl96093604,n457859}. Some of the possible schemes include but not limited to $\lambda$-type \cite{pra82032322}, $V$-type \cite{jpb42165506}, cascade-type \cite{oc2821593,jpb41035501}, and $Y$-type \cite{jpb43155506}. In these works, the coherent superposition is induced by exciting an atom in a lower energy level to the upper (using an external pumping mechanism) from where the atom undergoes  direct spontaneous emission that leads to a genuine tripartite entanglement \cite{epjd56247,oc2821593}. Quite recently, Shi {\it{et al.}} \cite{jpb43155506} extensively discussed the way of generating a tripartite entanglement applying a mechanism of six-wave mixing. The three external radiations they applied were believed to be responsible for creating the required nonclassical correlations. 

However, in this contribution, following a similar kind of reasoning,  it is proposed that the $Y$-shaped four-level atomic scheme can be a reliable source of strong entangled light if the initial preparation of the atoms is assumed to be accountable for inducing coherent superposition. Moreover, rather than placing the atoms in the cavity throughout the operation and expose them to thermal fluctuations, as usually the case, it is assumed that they are injected into the cavity at a constant rate. For the sake of convenience, the amplification of the light due to reflection between the walls of the coupler mirror when photons with different frequency emitted in the forked four-level cascade transition are correlated by the coherence induced via initial preparation is dubbed as a nondegenerate $Y$-shaped four-level correlated emission laser. Since a large number of atoms, in principle, can be injected into the cavity over a longer period of time and the direct spontaneous emission process can also be efficient, if the atoms are properly prepared initially, it would be reasonable expecting this system as a reliable source of bright light. 

In order to study the dynamics of the entanglement employing the existing criteria, it is found necessary and appropriate establishing the mathematical framework for solving the involved differential equations beforehand. To this effect, based on the involved structure of the atomic levels, the master equation is derived following the outline presented elsewhere for the corresponding two-mode case \cite{pra79033810,pra82053835}. Due to the coherent superposition induced via initial preparation and cascading process, a significant correlation in the generated three modes is observed in the calculated master equation. Moreover, with the intention of paving the way for in depth analysis, the procedure of solving the emerging coupled differential equations is outlined. It is found that the required solutions can be explicitly written down once the corresponding $3\times3$ matrix constructed from the prefactors in the master equation is diagonalized and its eigenmatrix is constructed, which in principle is a surmountable task although the rigor may be somewhat lengthy. 

\section{Description of the Model}

\begin{figure}[htb]
\centerline{\includegraphics [height=6.5cm,angle=0]{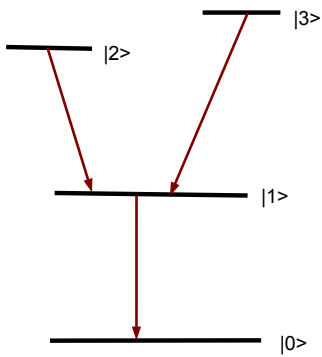}}
\caption {\label{fig1} Schematic representation of the nondegenerate $Y$-shaped four-level atom. It is assumed that the atom undergoes  direct spontaneous transitions from energy level $|3\rangle$ to  $|1\rangle$ in a process emits a photon describable by annihilation operator $\hat{a}_{3}$, when it goes from $|2\rangle$ to $|1\rangle$ $\hat{a}_{2}$, and when from $|1\rangle$ to $|0\rangle$ $\hat{a}_{1}$. The three transitions are presumed to be resonantly coupled to the cavity radiation.} \end{figure}

In earlier studies on the three-level cascade laser, it was observed that the spontaneous transition during the cascading process induces a coherent superposition that leads to enhanced quantum features including a bipartite CV entanglement \cite{pra75062305,oc283781}. It can be asserted that initially preparing atoms in a certain coherent superposition and then allowing them to follow realistic spontaneous transition roots can yield a strongly entangled light \cite{pra74043816,pra77013815,jpb42215506}. Taking this as a motivation, the $Y$-shaped four-level atomic system initially prepared in a coherent superpostion of the energy levels between which a direct electric dipole transition is forbidden would be considered. For the sake of convenience, the lower energy level is denoted by $|0\rangle$,  the intermediate energy level by $|1\rangle$, and the upper two energy levels by $|2\rangle$ and $|3\rangle$ (Please note that the schematic representation of the involved atomic energy levels is provided in Fig. \ref{fig1}). In order to expedite the cascading process, it is assumed that the parity of the energy levels $|0\rangle$, $|2\rangle$, and $|3\rangle$ is the same whereas that of $|1\rangle$ is different. This entails that  direct spontaneous transitions between energy levels $|2\rangle$ $\leftrightarrow$ $|3\rangle$, $|2\rangle$ $\leftrightarrow$ $|0\rangle$, and $|0\rangle$ $\leftrightarrow$ $|3\rangle$ are electric dipole forbidden, but due to the parity difference, the transitions between $|1\rangle$ $\leftrightarrow$ $|0\rangle$, $|2\rangle$ $\leftrightarrow$ $|1\rangle$, and $|1\rangle$ $\leftrightarrow$ $|3\rangle$ are allowed. It is worth noting that if required the dipole forbidden transitions can be induced by an external pumping mechanism in a similar manner as in the three-level case \cite{prl94023601, pra75033816,jpb41055503,jpb41145501}.

While the atom undergoes a direct spontaneous transition from energy level $|3\rangle$ to $|1\rangle$, suppose it emits a photon represented by annihilation operator $\hat{a}_{3}$. In principle, it can still undergo a direct spontaneous emission and go over to the lower energy level $|0\rangle$; in the process emits a photon described by $\hat{a}_{1}$. In the cascading transition from energy level $|3\rangle$ to $|0\rangle$ via  $|1\rangle$,  a correlation between the two emitted photons ($\hat{a}_{3}$ and $\hat{a}_{1}$) can readily be established. In a similar manner, it is not difficult to realize that there could be a correlation between the photons emitted ($\hat{a}_{2}$ and $\hat{a}_{1}$)  when the atom undergoes a spontaneous transition from energy level $|2\rangle$ to $|0\rangle$ via $|1\rangle$. These two processes, which are not entirely independent, are expected to initiate nonclassical correlations between the photons emitted while the atom cascades from the upper two energy levels to the lower via different forked roots. Nevertheless, to establish a genuine tripartite entanglement as prescribed by the von Loock and Furusawa criteria \cite{pra67052315}, a correlation between the photons emitted from the upper two energy levels ($\hat{a}_{2}$ and $\hat{a}_{3}$) is very crucial. In order to initiate this important correlation, it is worth noting that in case the atom is initially prepared in a coherent superposition of the upper two energy levels and the lower, it does not arbitrarily undergo the aforementioned spontaneous transitions due to the resulting population sharing. 

If there is a triply resonant light in the cavity, the atom with energy level $|1\rangle$ has three distinct alternatives to take except for the spontaneous decay to any other energy level that is not involved in the present consideration. The first and the most natural one is to continue with the direct spontaneous emission and then goes over to the lower energy level, or it can absorb a photon ($\hat{a}_{2}$) and excited to the upper energy level $|2\rangle$, or it can absorb a photon ($\hat{a}_{3}$) and excited to the upper energy level $|3\rangle$. In this description, if one manages to send a large number of initially prepared atoms through the cavity, the absorption-emission mechanism painstakingly follows different roots, which leads to additional correlation between emitted photons. In this scenario, as long as there is a coherent superposition between the energy levels $|2\rangle$ and $|3\rangle$ initially, a meaningful correlation in a subsequent emission of photons denoted by $\hat{a}_{2}$ and $\hat{a}_{3}$ is expected. It is, hence, envisaged that this process can forge the required nonclassical correlation between the photons emitted from the upper two energy levels. 

The more general explanation and possible setup for practical utilization of similar schemes were provided earlier in Refs. \cite{prl99123603,prl102013601}. However, this contribution has one essential difference in which the initial preparation is assumed to be a prominent source of the atomic coherent superposition as opposed to the external pumping mechanism. Although the initial preparation and injection process lead to some technical difficulties in practical utilization of the potential of this system, it is envisaged that the challenge would be less severe when compared to an external pumping mechanism that naturally initiates thermal fluctuations and atomic broadening.

\section{Master equation}

The interaction of a nondegenerate $Y$-shaped four-level atom with a triply resonant cavity
radiation can be described in the rotating-wave approximation and
the interaction picture by the Hamiltonian of the form
\begin{align}\label{fl1}\hat{H}_{I}&=ig\big[\hat{a}_{3}|3\rangle\langle1|-|1\rangle\langle3|\hat{a}^{\dagger}_{3}+\hat{a}_{2}|2\rangle\langle1|-|1\rangle\langle2|\hat{a}^{\dagger}_{2}\notag\\&+\hat{a}_{1}|1\rangle\langle0|-|0\rangle\langle1|\hat{a}^{\dagger}_{1}\big],\end{align} where $g$ is a coupling constant
chosen to be the same for all transitions for convenience and  $\hat{a}_{i}$'s are
the annihilation operators that represent the three cavity modes.

Assuming that the atoms are initially prepared in the coherent superposition of the atomic energy levels except the intermediate, the pertinent atomic state can be taken as
\begin{align}\label{fl2}|\Psi_{A}(0)\rangle=C_{3}(0)|3\rangle+C_{2}(0)|2\rangle+C_{0}(0)|0\rangle,\end{align}
where $C_{i}(0)$'s are the probability amplitudes for the atom to be initially in the $i$'s energy level and the corresponding density operator takes the form
\begin{align}\label{fl3}\rho_{A}^{(0)}&=\rho_{33}^{(0)}|3\rangle\langle3|+\rho_{32}^{(0)}|3\rangle\langle2|+\rho_{23}^{(0)}|2\rangle\langle3|+\rho^{(0)}_{30}|3\rangle\langle0|\notag\\&+\rho^{(0)}_{03}|0\rangle\langle3|+\rho^{(0)}_{22}|2\rangle\langle2|+\rho^{(0)}_{20}|2\rangle\langle0|+\rho^{(0)}_{02}|0\rangle\langle2|\notag\\&+\rho^{(0)}_{00}|0\rangle\langle0|,\end{align} where $\rho_{ii}^{(0)}$'s are the initial populations and $\rho_{ij(i\ne j)}^{(0)}$'s are the coherences between the atomic energy levels. It is worth noting that the intermediate energy level  is initially unpopulated and the phase fluctuation resulting from imperfect preparation is not taken into consideration.

 The atoms prepared in this manner are assumed to be injected into a triply resonant cavity at a constant rate $r_{a}$
and removed after sometime $T$, which is long enough for the
atoms to spontaneously decay to energy levels that do not involve in the process. The density operator
 for the cavity radiation plus a single atom  injected into the cavity at
time $t_{j}$ can be denoted by $\rho_{AR}(t,t_{j})$, where $t-T\le t_{j}\le
t$. The density operator for all the atoms in the cavity plus the
cavity radiation at time $t$ can be expressed as
\begin{align}\label{fl4}\hat{\rho}_{AR}(t) =
r_{a}\sum_{j}\hat{\rho}_{AR}(t,t_{j})\Delta t_{j},\end{align}
where $r_{a}\Delta t_{j}$ represents the number of atoms injected
into the cavity in a time interval of $\Delta t_{j}$. Assuming
that the atoms are continuously injected into the cavity and
taking the limit that $\Delta t_{j}\rightarrow 0$, the summation
over $j$ can be converted into integration with respect to $t'$,
\begin{align}\label{fl5}\hat{\rho}_{AR}(t) =
r_{a}\int_{t-T}^{t} \hat{\rho}_{AR}(t,t')dt'.\end{align} 
Replacing the summation over randomly
injected atoms to integration in a similar manner has been done frequently \cite{pr159208,pra40237,pra82053835}. 

It is also a well established fact
that the density operator  evolves in time
according to
\begin{align}\label{fl6}\frac{\partial}{\partial t}\hat{\rho}_{AR}(t,t') =
-i\big[\hat{H}, \;\hat{\rho}_{AR}(t,t')\big].\end{align}
 It is not hard to observe that $t'$ can be switched in such a way that $\hat{\rho}_{AR}(t,t)$ represents the density operator for an
 atom plus the cavity radiation at a time when the atom is injected into the
 cavity, whereas $\hat{\rho}_{AR}(t,t-T)$ represents the
 density operator
 when the atom is removed from the cavity. Since the atomic and radiation variables
 are not correlated at the instant the atoms
 are injected into or removed from  the cavity, it is possible to propose that
 \begin{align}\label{fl8}\hat{\rho}_{AR}(t,t) = \hat{\rho}_{A}(0)\hat{\rho}(t),\end{align}
 \begin{equation}\label{fl9}  \hat{\rho}_{AR}(t,t-T)=\hat{\rho}_{A}(t-T)\hat{\rho}(t),\end{equation}
 where $  \hat{\rho}_{A}(0)=\hat{\rho}_{A}(t).$

Hence, in view  of Eqs. \eqref{fl8} and \eqref{fl9}, integration of Eq. \eqref{fl5} results
\begin{equation}\label{fl13}\frac{d}{dt}\hat{\rho}_{AR}(t) =
r_{a}[\hat{\rho}_{A}(0) - \hat{\rho}_{A}(t-T)]\hat{\rho}(t) -
i[\hat{H}, \;\hat{\rho}_{AR}(t)].\end{equation} Now taking the trace over the atomic variables using the fact that
$Tr_{A}(\hat{\rho}_{A}(0)) = Tr_{A}(\hat{\rho}_{A}(t-T)) =1,$ leads to
\begin{align}\label{fl15}\frac{d\hat{\rho}(t)}{dt}
= - iTr_{A}[\hat{H}, \;\hat{\rho}_{AR}(t)].\end{align} 

Upon employing Eqs. \eqref{fl1} and \eqref{fl15}, the time evolution of the
reduced density operator for radiation turns out to be
\begin{align}\label{fl16}\frac{d\hat{\rho}(t)}{dt}& =
g\big[\hat{\rho}_{31}\hat{a}^{\dagger}_{3}-\hat{a}^{\dagger}_{3}\hat{\rho}_{31}+
 \hat{a}_{3}\hat{\rho}_{13} - \hat{\rho}_{13}\hat{a}_{3}
\notag\\&+\hat{\rho}_{21}\hat{a}^{\dagger}_{2}-\hat{a}^{\dagger}_{2}\hat{\rho}_{21}+
 \hat{a}_{2}\hat{\rho}_{12} - \hat{\rho}_{12}\hat{a}_{2}]\notag\\&+\hat{\rho}_{10}\hat{a}^{\dagger}_{1}-\hat{a}^{\dagger}_{1}\hat{\rho}_{10}+
 \hat{a}_{1}\hat{\rho}_{01} - \hat{\rho}_{01}\hat{a}_{1}\big],\end{align}
in which $\hat{\rho}_{ij} =\langle i|\hat{\rho}_{AR}|j\rangle$ with
$i,j =$ 0, 1, 2,  3.

On the other hand, on the basis of Eq. \eqref{fl13}, one can readily write
\begin{align}\label{fl18}\frac{d}{dt}\hat{\rho}_{ij}(t)& =
r_{a}\langle i|\hat{\rho}_{A}(0)|j\rangle\hat{\rho} -
r_{a}\langle i|\hat{\rho}_{A}(t-T)|j\rangle\hat{\rho} \notag\\&-
i\langle i|[\hat{H}, \;\hat{\rho}_{AR}(t)]|j\rangle -
\gamma\hat{\rho}_{ij},\end{align} where the last term is
introduced in order to account for the atomic decay process. Basically, $\gamma$ is the decay rate associated with every atomic transition including the rate of dephasing. In practical situation, assuming all decay rates as equal may not be reasonable as recently discussed elsewhere \cite{pra79033810,pra79063815}.

Assuming the atoms to be removed from
the cavity after they have decayed to energy levels that do not involve in the lasing process implies that $\langle i|\hat{\rho}_{A}(t-T)|j\rangle =0.$ Hence, on account of Eqs. \eqref{fl1}, \eqref{fl3}, and
\eqref{fl18}, one gets
\begin{align}\label{fl20}\frac{d}{dt}\hat{\rho}_{ij}(t) &=-\gamma\hat{\rho}_{ij}+
r_{a}\hat{\rho}(t)\big[\rho_{33}^{(0)}\delta_{i3}\delta_{3j} +
\rho_{32}^{(0)}\delta_{i3}\delta_{2j} \notag\\&+
\rho_{23}^{(0)}\delta_{i2}\delta_{3j} +
\rho_{30}^{(0)}\delta_{i3}\delta_{0j}+\rho_{03}^{(0)}\delta_{i0}\delta_{3j} +
\rho_{22}^{(0)}\delta_{i2}\delta_{2j} \notag\\&+
\rho_{20}^{(0)}\delta_{i2}\delta_{0j} +
\rho_{02}^{(0)}\delta_{i0}\delta_{2j}+\rho_{00}^{(0)}\delta_{i0}\delta_{0j}\big]
\notag\\&+ g[\hat{a}_{3}\hat{\rho}_{1j}\delta_{i3}-
\hat{a}^{\dagger}_{3}\hat{\rho}_{3j}\delta_{i1} +
\hat{a}_{2}\hat{\rho}_{1j}\delta_{i2} -
\hat{a}^{\dagger}_{2}\hat{\rho}_{2j}\delta_{i1} \notag\\&+ \hat{a}_{1}\hat{\rho}_{0j}\delta_{i1} -\hat{a}_{1}^{\dagger} \hat{\rho}_{1j}\delta_{0i}
-\hat{\rho}_{i3}\hat{a}_{3}\delta_{1j} +
\hat{\rho}_{i1}\hat{a}^{\dagger}_{3}\delta_{3j}  \notag\\&-
\hat{\rho}_{i2}\hat{a}_{2}\delta_{1j} +
\hat{\rho}_{i1}\hat{a}^{\dagger}_{2}\delta_{2j} -
\hat{\rho}_{i1}\hat{a}_{1}\delta_{0j} +
\hat{\rho}_{i0}\hat{a}^{\dagger}_{1}\delta_{1j}],\end{align} from which follows
\begin{align}\label{fl21}\frac{d}{dt}\hat{\rho}_{33}(t) &= r_{a}\rho_{33}^{(0)}\hat{\rho}(t)
+ g[\hat{a}_{3}\hat{\rho}_{13} + \hat{\rho}_{31}\hat{a}^{\dagger}_{3}]-\gamma\hat{\rho}_{33},\end{align}
\begin{align}\label{fl22}\frac{d}{dt}\hat{\rho}_{32}(t) &= r_{a}\rho_{32}^{(0)}\hat{\rho}(t)
+ g[\hat{a}_{3}\hat{\rho}_{12} + \hat{\rho}_{31}\hat{a}^{\dagger}_{2}]
-\gamma\hat{\rho}_{32},\end{align}
\begin{align}\label{fl27}\frac{d}{dt}\hat{\rho}_{31}(t) =g[\hat{a}_{3}\hat{\rho}_{11} - \hat{\rho}_{33}\hat{a}_{3}-
\hat{\rho}_{32}\hat{a}_{2} + \hat{\rho}_{30}\hat{a}^{\dagger}_{1}] -\gamma\hat{\rho}_{31},\end{align}
\begin{align}\label{fl23}\frac{d}{dt}\hat{\rho}_{30}(t) &= r_{a}\rho_{30}^{(0)}\hat{\rho}(t)
+ g[\hat{a}_{3}\hat{\rho}_{10} - \hat{\rho}_{31}\hat{a}_{1}]
-\gamma\hat{\rho}_{30},\end{align}
\begin{align}\label{fl24}\frac{d}{dt}\hat{\rho}_{22}(t) &= r_{a}\rho_{22}^{(0)}\hat{\rho}(t)
+ g[\hat{a}_{2}\hat{\rho}_{12} + \hat{\rho}_{21}\hat{a}^{\dagger}_{2}]-\gamma\hat{\rho}_{22},\end{align}
\begin{align}\label{fl28}\frac{d}{dt}\hat{\rho}_{21}(t) =g[\hat{a}_{2}\hat{\rho}_{11} - \hat{\rho}_{23}\hat{a}_{3}-
\hat{\rho}_{22}\hat{a}_{2} + \hat{\rho}_{20}\hat{a}^{\dagger}_{1}] -\gamma\hat{\rho}_{21},\end{align}
\begin{align}\label{fl25}\frac{d}{dt}\hat{\rho}_{20}(t) &= r_{a}\rho_{20}^{(0)}\hat{\rho}(t)
+ g[\hat{a}_{2}\hat{\rho}_{10} - \hat{\rho}_{21}\hat{a}_{1}]
-\gamma\hat{\rho}_{20},\end{align}
\begin{align}\label{fl29}\frac{d}{dt}\hat{\rho}_{11}(t)& =g[\hat{a}_{1}\hat{\rho}_{01} - \hat{\rho}_{13}\hat{a}_{3}-
\hat{\rho}_{12}\hat{a}_{2} + \hat{\rho}_{10}\hat{a}^{\dagger}_{1}\notag\\&-\hat{a}^{\dagger}_{3}\rho_{31}-\hat{a}^{\dagger}_{2}\rho_{21}] -\gamma\hat{\rho}_{11},\end{align}
\begin{align}\label{fl30}\frac{d}{dt}\hat{\rho}_{10}(t) =g[\hat{a}_{1}\hat{\rho}_{00} - \hat{\rho}_{11}\hat{a}_{1}-
-\hat{a}^{\dagger}_{3}\hat{\rho}_{30} - \hat{a}^{\dagger}_{2}\hat{\rho}_{20}] -\gamma\hat{\rho}_{10},\end{align}
\begin{align}\label{fl26}\frac{d}{dt}\hat{\rho}_{00}(t) &= r_{a}\rho_{00}^{(0)}\hat{\rho}(t)
- g[\hat{a}_{1}\hat{\rho}_{10} + \hat{\rho}_{01}\hat{a}^{\dagger}_{1}]
-\gamma\hat{\rho}_{00}.\end{align}

In the good cavity limit ($\gamma\gg\kappa$, where $\kappa$ is the cavity damping constant), the cavity mode
variables change slowly compared with the atomic variables. It is, hence, expected that
the atomic variables will reach steady state in a relatively short
time. The time derivative of such variables can be set to zero, while keeping the remaining atomic and cavity mode
variables at time $t$. This procedure is referred to as the
adiabatic approximation scheme. Confining to linear
analysis, which amounts to dropping the terms containing $g$ in
Eqs. \eqref{fl21}, \eqref{fl22}, \eqref{fl23}, \eqref{fl24}, \eqref{fl25}, and \eqref{fl26},
and applying the adiabatic approximation scheme, one finds
\begin{align}\label{fl31}\hat{\rho}_{ij}= {r_{a}\rho_{ij}^{(0)}\over\gamma}\hat{\rho}(t),\end{align}
\begin{align}\label{fl32}\hat{\rho}_{11} =0,\end{align}
with $\hat{\rho} = \hat{\rho}(t)$ and $i,j=0,2,3$. 
Eq. \eqref{fl32} indicates the procedure in which the initially unpopulated energy level ($|1\rangle$) is adiabatically eliminated.

At this juncture, making use of Eqs. \eqref{fl27}, \eqref{fl28}, \eqref{fl29}, \eqref{fl30},
\eqref{fl31}, \eqref{fl32},
and  applying the adiabatic
approximation scheme once again, it is possible to verify that
\begin{align}\label{fl36}\hat{\rho}_{31}& =\frac{gr_{a}\hat{\rho}}{\gamma^{2}}\left[\rho_{30}^{(0)}\hat{a}_{1}^{\dagger} -
\rho_{33}^{(0)}\hat{a}_{3} -\rho_{32}^{(0)}\hat{a}_{2}\right],\end{align}
\begin{align}\label{fl37}\hat{\rho}_{21}& =\frac{gr_{a}\hat{\rho}}{\gamma^{2}}\left[\rho_{20}^{(0)}\hat{a}_{1}^{\dagger} -
\rho_{22}^{(0)}\hat{a}_{2} -\rho_{32}^{(0)}\hat{a}_{3}\right],\end{align}
\begin{align}\label{fl38}\hat{\rho}_{10}& =\frac{gr_{a}}{\gamma^{2}}\left[\rho_{00}^{(0)}\hat{a}_{1} -
\rho_{20}^{(0)}\hat{a}_{2}^{\dagger} -\rho_{30}^{(0)}\hat{a}_{3}^{\dagger}\right]\hat{\rho},\end{align} where $\rho_{ij}^{(0)}=\rho_{ji}^{(0)}$ is set based on the fact that the initial populations and coherences are real constants.

Now applying
Eqs. \eqref{fl36}, \eqref{fl37}, and \eqref{fl38}, it is possible to express Eq. \eqref{fl16} as
\begin{widetext}\begin{align}\label{fl39}\frac{d\hat{\rho}}{dt}& =\frac{r_{a}g^{2}\rho_{33}^{(0)}}{\gamma^{2}} \left[2\hat{a}^{\dagger}_{3}\hat{\rho}\hat{a}_{3} -
 \hat{\rho}\hat{a}_{3}\hat{a}^{\dagger}_{3} - \hat{a}_{3}\hat{a}^{\dagger}_{3}\hat{\rho}\right] +
\frac{r_{a}g^{2}\rho_{32}^{(0)}}{\gamma^{2}} \left[2\hat{a}^{\dagger}_{3}\hat{\rho}\hat{a}_{2}+2\hat{a}_{2}^{\dagger}\hat{\rho}\hat{a}_{3} -
\hat{\rho}\hat{a}_{2}\hat{a}^{\dagger}_{3} - \hat{a}_{3}\hat{a}^{\dagger}_{2}\hat{\rho}-\hat{\rho}\hat{a}_{3}\hat{a}_{2}^{\dagger}-\hat{a}_{2}\hat{a}_{3}^{\dagger}\hat{\rho}\right]\notag\\&+
\frac{r_{a}g^{2}\rho_{22}^{(0)}}{\gamma^{2}} \left[2\hat{a}^{\dagger}_{2}\hat{\rho}\hat{a}_{2} -
 \hat{\rho}\hat{a}_{2}\hat{a}^{\dagger}_{2} - \hat{a}_{2}\hat{a}^{\dagger}_{2}\hat{\rho}\right]-\frac{r_{a}g^{2}\rho_{30}^{(0)}}{\gamma^{2}} \left[2\hat{a}_{1}\hat{\rho}\hat{a}_{3}+2\hat{a}_{3}^{\dagger}\hat{\rho}\hat{a}_{1}^{\dagger} -
\hat{\rho}\hat{a}_{1}^{\dagger}\hat{a}^{\dagger}_{3} - \hat{a}_{1}^{\dagger}\hat{a}^{\dagger}_{3}\hat{\rho}-\hat{a}_{3}\hat{a}_{1}\hat{\rho}-\hat{\rho}\hat{a}_{3}\hat{a}_{1}\right]\notag\\&+
\frac{r_{a}g^{2}\rho_{00}^{(0)}}{\gamma^{2}} \left[2\hat{a}_{1}\hat{\rho}\hat{a}_{1}^{\dagger} -
 \hat{\rho}\hat{a}_{1}^{\dagger}\hat{a}_{1} - \hat{a}_{1}^{\dagger}\hat{a}_{1}\hat{\rho}\right]
-\frac{r_{a}g^{2}\rho_{20}^{(0)}}{\gamma^{2}} \left[2\hat{a}_{1}\hat{\rho}\hat{a}_{2}+2\hat{a}_{2}^{\dagger}\hat{\rho}\hat{a}_{1}^{\dagger} -
\hat{\rho}\hat{a}_{2}\hat{a}_{1} - \hat{a}_{2}\hat{a}_{1}\hat{\rho}-\hat{a}_{1}^{\dagger}\hat{a}_{2}^{\dagger}\hat{\rho}-\hat{\rho}\hat{a}_{1}^{\dagger}\hat{a}_{2}^{\dagger}\right].\end{align}\end{widetext} 

It is not difficult to observe from the form of this master equation that the two roots of the spontaneous transition resemble their counterparts in the three-level cascade scheme \cite{pra74043816}. It is good to note that, in addition to the correlation pertinent to the cascading process, the initial preparation also contributes a new term with a prefactor ${r_{a}g^{2}\rho_{32}^{(0)}\over\gamma^{2}}$ since $\rho_{32}^{(0)}$ represents the initial coherence associated with these atomic energy levels. Based on the fact that the cross-correlation terms are indicative of nonclassical features, it is possible to observe that this master equation can be taken as evidence for the existence of correlation between the three emitted photons.

In Eq. \eqref{fl39} there are six different prefactors that are not independent altogether, which might make the analysis more difficult. As a result, in order to rewrite this master equation in a more appealing manner, it appears convenient introducing two parameters defined by
\begin{align}\label{fl43}\eta_{1}=\rho_{00}^{(0)}-\rho_{33}^{(0)},\end{align}
\begin{align}\label{fl44}\eta_{2}=\rho_{00}^{(0)}-\rho_{22}^{(0)}.\end{align}  $\eta_{1}$ and $\eta_{2}$ are basically the population inversions viewed from different upper energy levels. In light of the anticipated initial preparation, $\eta_{1}$ and $\eta_{2}$ are not entirely independent. This can be evinced by the fact that \newline$\rho_{33}^{(0)}+\rho_{22}^{(0)}+\rho_{00}^{(0)}=1$, which leads to
\begin{align}\label{fl45}\rho_{00}^{(0)}={1+\eta_{1}+\eta_{2}\over3},\end{align}
\begin{align}\label{fl46}\rho_{22}^{(0)}={1+\eta_{1}-2\eta_{2}\over3},\end{align}
\begin{align}\label{fl47}\rho_{33}^{(0)}={1+\eta_{2}-2\eta_{1}\over3}.\end{align} 
Moreover, based on the nature of the initial state (Eq. \eqref{fl2}), that is, $\rho_{33}^{(0)}=C_{3}(0)C^{*}_{3}(0)$, $\rho_{22}^{(0)}=C_{2}(0)C^{*}_{2}(0)$, and \newline$\rho_{00}^{(0)}=C_{0}(0)C^{*}_{0}(0)$, it may not be difficult to realize that
$\rho_{30}^{(0)}=\sqrt{\rho_{33}^{(0)}\rho_{00}^{(0)}},$
$\rho_{20}^{(0)}=\sqrt{\rho_{22}^{(0)}\rho_{00}^{(0)}},$ and
$\rho_{32}^{(0)}=\sqrt{\rho_{33}^{(0)}\rho_{22}^{(0)}}.$

Furthermore, assuming that the environment modes can be represented by a three-mode independent vacuum reservoir, it is possible to include its effect following the standard approach \cite{lou}. In this respect, with the aid of the above variable transformation, one readily finds
\begin{widetext}\begin{align}\label{fl48}\frac{d\hat{\rho}}{dt}& =AB \left[2\hat{a}^{\dagger}_{3}\hat{\rho}\hat{a}_{3} -
 \hat{\rho}\hat{a}_{3}\hat{a}^{\dagger}_{3} - \hat{a}_{3}\hat{a}^{\dagger}_{3}\hat{\rho}\right]+
AE \left[2\hat{a}^{\dagger}_{3}\hat{\rho}\hat{a}_{2}-
\hat{\rho}\hat{a}_{2}\hat{a}^{\dagger}_{3}-\hat{a}_{2}\hat{a}_{3}^{\dagger}\hat{\rho}+2\hat{a}_{2}^{\dagger}\hat{\rho}\hat{a}_{3} -\hat{\rho}\hat{a}_{3}\hat{a}_{2}^{\dagger} - \hat{a}_{3}\hat{a}^{\dagger}_{2}\hat{\rho}\right] \notag\\&+
AC \left[2\hat{a}^{\dagger}_{2}\hat{\rho}\hat{a}_{2} -
 \hat{\rho}\hat{a}_{2}\hat{a}^{\dagger}_{2} - \hat{a}_{2}\hat{a}^{\dagger}_{2}\hat{\rho}\right]-
AF \left[2\hat{a}_{1}\hat{\rho}\hat{a}_{3}-\hat{a}_{3}\hat{a}_{1}\hat{\rho}-\hat{\rho}\hat{a}_{3}\hat{a}_{1}+2\hat{a}_{3}^{\dagger}\hat{\rho}\hat{a}_{1}^{\dagger} -
\hat{\rho}\hat{a}_{1}^{\dagger}\hat{a}^{\dagger}_{3} - \hat{a}_{1}^{\dagger}\hat{a}^{\dagger}_{3}\hat{\rho}\right]\notag\\&+
AD \left[2\hat{a}_{1}\hat{\rho}\hat{a}_{1}^{\dagger} -
 \hat{\rho}\hat{a}_{1}^{\dagger}\hat{a}_{1} - \hat{a}_{1}^{\dagger}\hat{a}_{1}\hat{\rho}\right]
-AG \left[2\hat{a}_{1}\hat{\rho}\hat{a}_{2}-
\hat{\rho}\hat{a}_{2}\hat{a}_{1} - \hat{a}_{2}\hat{a}_{1}\hat{\rho}+2\hat{a}_{2}^{\dagger}\hat{\rho}\hat{a}_{1}^{\dagger} -\hat{a}_{1}^{\dagger}\hat{a}_{2}^{\dagger}\hat{\rho}-\hat{\rho}\hat{a}_{1}^{\dagger}\hat{a}_{2}^{\dagger}\right]\notag\\&+{\kappa\over2}\sum_{i=1}^{3}\left[2\hat{a}_{i}\hat{\rho}\hat{a}^{\dagger}_{i}-\hat{a}^{\dagger}_{i}\hat{a}_{i}\hat{\rho}-\hat{\rho}\hat{a}_{i}^{\dagger}\hat{a}_{i}\right],\end{align}\end{widetext} where
\begin{align}\label{fl49}A = \frac{2r_{a}g^{2}}{\gamma^{2}},\end{align}
\begin{align}\label{fl50}B=\frac{1+\eta_{2}-2\eta_{1}}{6},\end{align}
\begin{align}\label{fl51}C=\frac{1+\eta_{1}-2\eta_{2}}{6},\end{align}
\begin{align}\label{fl52}D=\frac{1+\eta_{1}+\eta_{2}}{6},\end{align}
\begin{align}\label{fl53}E=\frac{\sqrt{1-\eta_{1}-\eta_{2}+5\eta_{1}\eta_{2}-2(\eta_{1}^{2}+\eta^{2}_{2})}}{6},\end{align}
\begin{align}\label{fl54}F=\frac{\sqrt{1-\eta_{1}+2\eta_{2}-\eta_{1}\eta_{2}-2\eta_{1}^{2}+\eta^{2}_{2}}}{6},\end{align}
\begin{align}\label{fl55}G=\frac{\sqrt{1-\eta_{2}+2\eta_{1}-\eta_{1}\eta_{2}+\eta_{1}^{2}-2\eta^{2}_{2}}}{6},\end{align}
and $\kappa$ is the cavity damping constant taken to be the same for all modes for convenience. 

\section{Demonstration of the tripartite entanglement}

It is clearly shown that this master equation is essentially described in terms of $A$, $\eta_{1}$, $\eta_{2}$, and $\kappa$. It is possible to infer from the form of the master equation that the first two terms to the left indicate the gain in modes $\hat{a}_{3}$ and $\hat{a}_{2}$, in respective order, whereas the third term the loss of mode $\hat{a}_{1}$. This outcome is fairly consistent with the earlier reports on the cascade three-level scheme. As already discussed previously, there is a cross-correlation between the three modes, primarily in the form of $\hat{a}_{1}$ and $\hat{a}_{2}$, $\hat{a}_{1}$ and $\hat{a}_{3}$, and  $\hat{a}_{2}$ and $\hat{a}_{3}$, separately. Hence from the outset, it is not hard to envisage, according to the von Loock and Furusawa criteria \cite{pra67052315}, a genuine CV tripartite entanglement of a radiation generated by a coherently prepared Y-shaped four-level laser. It is obvious that the detail of the detectable degree of entanglement depends on the strength of $E$, $F$, and $G$, which on the other hand heavily rely on $A$, $\eta_{1}$, and $\eta_{2}$. It goes without saying that the strength of the entanglement by and large depends on the way the atoms are initially prepared and the rate at which the atoms are injected, which believed to give the experimenter a considerable freedom for manipulation.

It is good to note that the relation between $\eta_{1}$ and $\eta_{2}$ is so subtle that it can lead to very rich alternatives in analyzing the system. For instance, when the atoms are initially prepared to be in the lower energy level, one can readily see that $\eta_{1}=\eta_{2}=1$. In this case, $B=C=0$, which indicates that no photon is generated from the two upper levels. Moreover, $E=F=G=0,$ which implies that virtually there is no anticipated correlation; as it should be. However, assuming the atoms to be prepared initially in an equal probability between the three levels, $\eta_{1}=\eta_{2}=0,$  leads to $B=C=D=E=F=G=1/6$. This indicates that there is a meaningful correlation among the emitted radiations. Basically, these two options are the two extreme cases where there is no  and possible maximum coherence at the beginning, respectively. 

For the sake of convenience, suppose the atoms are initially prepared so that 50\% of them to be in the lower energy level, that is, $\eta_{1}+\eta_{2}=0.5$ and the remaining 50\% of them are in one of the upper  energy levels (let us say $|3\rangle$); $\eta_{1}=0$ and $\eta_{2}=0.5$. In this case, one can readily see that $C=F=G=0$, which shows that there is only one part of the cascade transitions. Since there is no photon with $\hat{a}_{2}$,  tripartite entanglement is not expected. In order to see the situation in depth, with the same assumption regarding to the lower energy level, suppose the remaining 50\% population is equally shared between the upper two energy levels, that is, $\eta_{1}=\eta_{2}=0.25$. In this case, it is not difficult to assert that all prefactors in the master equation are different from zero, which indicates the possibility for having nonclassical correlations among the emitted photons. It is not difficult to observe, at this juncture, that a similar outcome could have been predicted had the 50\% population is arbitrarily shared between the upper two energy levels; although the details can vary.

With the same convection, suppose the atoms are initially prepared in 50:50 coherent superposition of the upper  two energy levels; $\eta_{1}=\eta_{2}=-0.5$ (please note that in this case the lower energy level is initially unpopulated). In this case, one can readily see that  $D=F=G=0$. This indicates that the radiation emitted in the spontaneous transition between the lower two energy levels is not correlated with the corresponding upper two transitions. This can be directly linked to the fact that since the upper two energy levels are prepared with a maximum coherent superposition between them, the atomic transition is basically restricted to  transitions from energy level $|3\rangle$ to $|1\rangle$ and then to $|2\rangle$ and vice versa. The chance that the atoms break the established coherence and goes over to the lower energy level is quite small. That is why even the corresponding mean photon number of the radiation which largely depends on a prefactor $D$ also can be quite small. 
In the same manner, it is possible to assert that there could be a meaningful correlation between the photons emitted during a direct spontaneous transition from the upper two energy levels to the lower due to the coherence induced by the cascading process  when the initial coherent superposition is not the maximum possible. The nonclassical correlation in this system seems to show the reminiscent of the absence of the bipartite entanglement in a three-level cascade system when the lower and the upper energy levels are prepared in a maximum coherent superposition \cite{jpb42215506}. 

One may deduce from this interpretation that the nonclassical correlation that leads to a tripartite entanglement can be induced by initially preparing atoms coherently between the upper two energy levels and the lower with arbitrary, other than zero, probability. It is also envisioned that the cascading transitions are a very vital mechanism in connecting the upper two energy levels with the lower. Therefore, no doubt that, properly harnessing the utility accorded with the initial preparation and cascading mechanisms results a genuine tripartite entangled light. Based on earlier studies in the three-level cascade laser \cite{jpb41055503,jpb41145501}, it is  equally expected that externally pumping the atoms can establish a coupling between energy levels in which direct spontaneous transitions are dipole forbidden that can significantly improve the tripartite nonclassical correlations.

\section{Conclusion}

The detailed derivation of the master equation that describes the radiation emitted from a coherently prepared nondegenerate $Y$-shaped four-level correlated emission laser is presented. In view of limiting the otherwise arising complications, the atoms are presumed to be prepared in an arbitrary perfect coherent superposition of the three energy levels, where the intermediate energy level is taken to be unpopulated at the beginning to make use of the adiabatic elimination technique. In setting up the laser, the initially prepared atoms are presumed to be injected into a triply resonant cavity. To pave the way for further in depth analysis, the approaches by which the corresponding stochastic differential equations can be obtained and the resulting equations are solved are outlined in the Appendix. Moreover, the rate equations that describe the time evolution of various correlations that can be required in the study of the quantum features and statistical properties of the radiation and the procedure applied in solving them are provided. 

It turns out that the quantum system under consideration can be a source of a continuous variable tripartite entangled light under certain conditions. Detailed investigation shows that due to the cascade transitions, the emission-absorption mechanism which is guided by the induced coherent superposition is found to be responsible for establishing the required correlation between the emitted photons. Further analysis based on varying the way the atoms are initially prepared shows that coherently coupling the three atomic energy levels is very vital in generating a tripartite entangled light. It is unequivocally asserted that leaving one of the upper energy levels unpopulated at the beginning leads to a bipartite entanglement at best since there is no way for the atoms to go to the unpopulated level in the course of the process. However, initially preparing the atoms in the upper two energy levels other than in a possible maximum coherence leaving the lower energy level unpopulated can lead to the appearance of the tripartite entanglement since the upper two energy levels can be coupled to the lower via the cascade transitions.

In relation to the similarity of the result of preparing the atoms in the coherent superposition of the upper two energy levels with the corresponding three-level scheme, it is expected that an external driving mechanism can be used to improve the generated entanglement in some respect. Furthermore, it may not be hard to realize that a highly intense light can be generated since the injection mechanism allows to send as many as required atoms through the cavity over a longer period of time without significantly exposing them to fluctuations and broadening associated with heating. Hence, this study by and large tries to show that the nondegenerate $Y$-shaped four-level scheme can be a source of reliable bright genuine tripartite entanglement with much more promise of flexible arrangement essentially by combining the initial preparation and external driving options. Even though starting with the form of the master equation and the values of the involved prefactors yield encouraging outcomes, it is incontestable that in depth analysis is still lacking. Owing to the involved rigor, length of the expressions, and complications of the different viable scenarios, a simpler and more specialized analysis is deferred to subsequent communications.

\section*{ACKNOWLEDGMENTS}

I thank the Max Planck Institute for the Physics of  Complex Systems for allowing me to visit and use their facility in carrying out this research and Dilla University for granting the leave of absence. I also acknowledge the valuable comments of Klaus Hornberger.

\section*{APPENDIX}

In order to study the quantum features and statistical properties of the radiation, different auto-correlations, that refer to the pertinent mean photon numbers, and cross-correlations are required. In many instances, it is possible to obtain these correlations by making use of  the master equation and then solve the resulting coupled differential equations. To this
effect, employing Eq. \eqref{fl48} and the fact that
${d\over dt}\langle\hat{O}(t)\rangle =
Tr\left({d\hat{\rho}\over dt}\hat{O}\right),$ where $\hat{O}$ is an operator, the following essential correlations are obtained (Please note that the full list is unnecessarily long)
$${d\over dt}\langle\hat{a}^{\dagger}_{1}(t)\rangle =-\left({\kappa\over2}+AD\right)\langle\hat{a}^{\dagger}_{1}(t)\rangle$$$$+AF\langle\hat{a}_{3}(t)\rangle+AG\langle\hat{a}_{2}(t)\rangle,$$
$${d\over dt}\langle\hat{a}_{2}(t)\rangle =-\left({\kappa\over2}-AC\right)\langle\hat{a}_{2}(t)\rangle$$$$+AE\langle\hat{a}_{3}(t)\rangle-AG\langle\hat{a}_{1}^{\dagger}(t)\rangle,$$
$${d\over dt}\langle\hat{a}_{3}(t)\rangle =-\left({\kappa\over2}-AB\right)\langle\hat{a}_{3}(t)\rangle$$$$+AE\langle\hat{a}_{2}(t)\rangle-AF\langle\hat{a}_{1}^{\dagger}(t)\rangle,$$
$${d\over dt}\langle\hat{a}_{3}^{\dagger}\hat{a}_{3}\rangle= (2AB-\kappa)\langle\hat{a}_{3}^{\dagger}\hat{a}_{3}\rangle +AE[\langle\hat{a}_{3}^{\dagger}\hat{a}_{2}\rangle+\langle\hat{a}_{3}\hat{a}_{2}^{\dagger}\rangle]$$$$- AF[\langle\hat{a}_{3}^{\dagger}\hat{a}_{1}^{\dagger}\rangle+\langle\hat{a}_{3}\hat{a}_{1}\rangle]+2AB,$$
$${d\over dt}\langle\hat{a}_{2}^{\dagger}\hat{a}_{2}\rangle= (2AC-\kappa)\langle\hat{a}_{2}^{\dagger}\hat{a}_{2}\rangle +AE[\langle\hat{a}_{3}^{\dagger}\hat{a}_{2}\rangle+\langle\hat{a}_{3}\hat{a}_{2}^{\dagger}\rangle]$$$$- AG[\langle\hat{a}_{2}^{\dagger}\hat{a}_{1}^{\dagger}\rangle+\langle\hat{a}_{2}\hat{a}_{1}\rangle]+2AC,$$
$${d\over dt}\langle\hat{a}_{1}^{\dagger}\hat{a}_{1}\rangle=- (2AD+\kappa)\langle\hat{a}_{1}^{\dagger}\hat{a}_{1}\rangle +AF[\langle\hat{a}_{3}^{\dagger}\hat{a}_{1}^{\dagger}\rangle+\langle\hat{a}_{3}\hat{a}_{1}\rangle]$$$$+ AG[\langle\hat{a}_{1}^{\dagger}\hat{a}_{2}^{\dagger}\rangle+\langle\hat{a}_{2}\hat{a}_{1}\rangle]-2AD,$$
$${d\over dt}\langle\hat{a}_{3}^{\dagger}\hat{a}_{2}\rangle= (AB+AC-\kappa)\langle\hat{a}_{3}^{\dagger}\hat{a}_{2}\rangle-AG\langle\hat{a}^{\dagger}_{3}\hat{a}_{1}^{\dagger}\rangle$$$$+ AE[\langle\hat{a}_{3}^{\dagger}\hat{a}_{3}\rangle+\langle\hat{a}_{2}^{\dagger}\hat{a}_{2}\rangle] -AF\langle\hat{a}_{2}\hat{a}_{1}\rangle+2AE,$$
$${d\over dt}\langle\hat{a}_{3}\hat{a}_{1}\rangle= (AB-AD-\kappa)\langle\hat{a}_{3}\hat{a}_{1}\rangle+ AG\langle\hat{a}_{3}\hat{a}_{2}^{\dagger}\rangle$$$$+AF[\langle\hat{a}_{3}^{\dagger}\hat{a}_{3}\rangle-\langle\hat{a}_{1}^{\dagger}\hat{a}_{1}\rangle]+AF,$$
$${d\over dt}\langle\hat{a}_{2}\hat{a}_{1}\rangle= (AC-AD-\kappa)\langle\hat{a}_{2}\hat{a}_{1}\rangle+ AE\langle\hat{a}_{1}\hat{a}_{3}\rangle $$$$+AF\langle\hat{a}_{2}\hat{a}_{3}^{\dagger}\rangle+AG[\langle\hat{a}_{2}^{\dagger}\hat{a}_{2}\rangle-\langle\hat{a}_{1}^{\dagger}\hat{a}_{1}\rangle]+AG.\eqno(A1)$$

Assuming that $c$-number expressions associated with normal ordering are mathematically more appealing, the first three equations are rewritten as
$${d\over dt}\alpha^{*}_{1}(t) =-\left({\kappa\over2}+AD\right)\alpha^{*}_{1}(t) +AF\alpha_{3}(t)$$$$+AG\alpha_{2}(t)+f^{*}_{1}(t),$$
$${d\over dt}\alpha_{2}(t) =-\left({\kappa\over2}-AC\right)\alpha_{2}(t) +AE\alpha_{3}(t)$$$$-AG\alpha_{1}^{*}(t)+f_{2}(t),$$
$${d\over dt}\alpha_{3}(t) =-\left({\kappa\over2}-AB\right)\alpha_{3}(t) +AE\alpha_{2}(t)$$$$-AF\alpha_{1}^{*}(t)+f_{3}(t),\eqno(A2)$$
where $f_{i}(t)$'s are the pertinent stochastic noise forces. 

For the sake of convenience, these equations can be put in a more compact from as
\begin{equation*}{d\over dt}{\cal{R}}(t)=-{\cal{M}}{\cal{R}}(t)+{\cal{N}}(t),\eqno(A3)\end{equation*} where
\begin{equation*}{\cal{M}}=\left(\begin{matrix}
  {\kappa\over2}+AD &-AG  & -AF \\
  AG & {\kappa\over2}-AC & -AE \\
  AF & -AE & {\kappa\over2}-AB
\end{matrix}\right),\end{equation*}
\begin{equation*}{\cal{R}}(t)=\left(\begin{matrix}
\alpha_{1}^{*}(t)  \\
  \alpha_{2}(t)\\
 \alpha_{3}(t)
                                    \end{matrix}\right),\end{equation*}
\begin{equation*}{\cal{N}}(t)=\left(\begin{matrix}
f_{1}^{*}(t)  \\
 f_{2}(t)\\
 f_{3}(t)
                                    \end{matrix}\right).\eqno(A4)\end{equation*} 

In principle, this coupled differential equations can be solved following a somewhat lengthy but straightforward algebra. First of all, it is desirable and possible to diagonalize the matrix ${\cal{M}}$ using the eigenvalue equations in which ${\cal{M}} {\cal{V}}_{i}=\lambda_{i}{\cal{V}}_{i},$
where ${\cal{V}}_{i}$'s are the eigenvectors and $\lambda_{i}$'s are the corresponding eigenvalues. For a 3X3 matrix, although the involved rigor is lengthy, it is possible to find both the eigenvalues and the corresponding eigenvectors, with the property that ${\cal{V}} {\cal{V}}^{-1}=\cal{I}$ and ${\cal{D}}={\cal{V}}^{-1} {\cal{M}} {\cal{V}},$ where ${\cal{V}}^{-1}$ is the inverse of the matrix constructed from relevant eigenvalues and ${\cal{D}}$ is the diagonal matrix corresponding to ${\cal{M}}$. With this arrangement, the solution of Eq. (A3) can be proposed as
\begin{equation*} {\cal{R}}(t)=\big[{\cal{V}} e^{-{\cal{D}} t}{\cal{V}}^{-1}\big] {\cal{R}}(0) +\int_{0}^{t}\big[ {\cal{V}} e^{-{\cal{D}}(t-t')}{\cal{V}}^{-1}\big] {\cal{N}}(t')dt'.\eqno(A5)\end{equation*}

It is common knowledge that ${\cal{R}}(0)$ describes the properties at the beginning of the lasing process. If initially the cavity is assumed to be in a three-mode vacuum state, one can readily disregard the contribution of the first term in Eq. (A5) and it is also possible to verify by taking the expectation values of the expressions in Eq. (A2) and comparing them with the first three expressions in Eq. (A1) that
\begin{equation*}\langle{\cal{N}}(t)\rangle=0,\eqno(A6)\end{equation*} which implies that $\langle f_{i}(t)\rangle=0$ for $i=1,2,3$. This essentially reflects the stochastic nature of the noise.

Furthermore, applying the various rate equations, distinct terms of Eq. (A5), and  assuming the noise force at later time does not affect system variables at earlier time result in 
\begin{equation*}{\cal{F}}(t',t'')=A\left(\begin{matrix}
 -2D & G & F \\
  G& 2C &2E  \\
  F& 2E & 2B
                                  \end{matrix}\right)\delta(t'-t''),\eqno(A7)\end{equation*} where 
${\cal{F}}(t,t')=\langle{\cal{N}}(t) {\cal{N}}^{T}(t')\rangle,$ in which $T$ stands for complex conjugate transpose. Based on the results obtained so far, it can be observed that
\begin{equation*}\langle{\cal{R}}(t){\cal{R}}^{T}(t)\rangle=\int_{0}^{t}\int_{0}^{t}{\cal{P}}(t,t'){\cal{F}}(t',t''){\cal{P}}^{T}(t,t'')dt'dt'',\eqno(A8)\end{equation*}
where
${\cal{P}}(t,t_{i})={\cal{V}}e^{{\cal{D}}(t-t_{i})}{\cal{V}}^{-1}.$

In principle, carrying out the involved matrix manipulations and then term by term integrations yield the correlations required for studying the quantum features and statistical properties of the radiation without further approximation. Even though the number of terms to be handled is somewhat large, carrying out the integration is over simplified by the presence of the $\delta$-function associated with the correlation of the noise forces (${\cal{F}}(t',t'')$) and the exponential dependence.

\end{document}